\documentclass[fleqn,usenatbib,useAMS]{mnras}

\usepackage[usenames,dvipsnames]{xcolor}
\usepackage{graphicx}	
\usepackage{amsmath}	
\usepackage{amssymb}	
\usepackage{multicol}        
\usepackage{multirow}
\usepackage{graphicx}
\usepackage{bm}		
\usepackage{pdflscape}	
\usepackage[T1]{fontenc}
\usepackage{tgbonum}
\usepackage{float}
\usepackage{siunitx}
\usepackage{appendix}

\graphicspath{{../figures/}}

\newcommand{\kms}{\,km\,s$^{-1}$} 

\usepackage[T1]{fontenc}
\usepackage{ae,aecompl}

\usepackage{newtxtext,newtxmath}

\title[AGN-Merger Connection]{ Tracing the AGN-Merger Connection: Insights from Cosmological Simulations and JWST Mock Observations}

\author[Jhee et al.]{Hannah Jhee$^{1}$,
Ena Choi$^{1}$\thanks{E-mail: enachoi@uos.ac.kr},
Rachel S. Somerville$^{2}$,
Dale D. Kocevski$^{3}$,
Michaela Hirschmann$^{4,5}$,
\newauthor{Thorsten Naab$^{6}$,
Desika Narayanan$^{7,8}$,
Intae Jung$^{9}$,
Juhan Kim$^{10}$}
\\
$^{1}$Department of Physics, University of Seoul, 163 Seoulsiripdaero, Dongdaemun-gu, Seoul 02504, Republic of Korea\\
$^{2}$Center for Computational Astrophysics, Flatiron Institute, 162 5th Ave., 
    New York, NY 10010, USA \\
$^{3}$Department of Physics and Astronomy, Colby College, Waterville, ME 04961, USA\\
$^{4}$Institut de Physique, Laboratoire d'astrophysique, \'Ecole Polytechnique F\'ed\'erale de Lausanne (EPFL), CH-1290 Versoix, Switzerland\\
$^{5}$DARK, Niels Bohr Institute, University of Copenhagen, Lyngbyvej 2, DK-2100 Copenhagen, Denmark\\
$^{6}$Max-Planck-Institut f{\"u}r Astrophysik, Karl-Schwarzschild Stra{\ss}e 1, 85748 Garching, Germany\\
$^{7}$Department of Astronomy, University of Florida, 211 Bryant Space Sciences centre, Gainesville, FL 32611, USA\\
$^{8}$University of Florida Informatics Institute, 432 Newell Drive, CISE Bldg E251, Gainesville, FL 32611, USA\\
$^{9}$Space Telescope Science Institute, 3700 San Martin Drive, Baltimore, MD 21218, USA\\
$^{10}$Korea Institute for Advanced Study (KIAS), 85 Hoegiro, Dongdaemun-gu, Seoul 02455, Republic of Korea
}

\date{Last updated 2025 May 21; in original form 2025 May 21}

\pubyear{2025}

\begin{document}
\label{firstpage}
\pagerange{\pageref{firstpage}--\pageref{lastpage}}
\maketitle

\begin{abstract}
Galaxy mergers have long been proposed as a mechanism for funneling gas toward galactic centres, potentially triggering accretion onto supermassive black holes (SMBHs) and igniting active galactic nuclei (AGN). While simulations often support this scenario, observational studies have yielded conflicting results regarding the AGN-merger connection.
In this study, we analyze 31 galaxies from cosmological zoom-in simulations spanning redshifts $0.5 < z < 3$. We identify mergers using detailed merger trees based on six-dimensional dark matter particle information and identify AGN activity through SMBH accretion histories.
To bridge the gap between simulations and observations, we generate mock JWST-like images and extract non-parametric morphological parameters. 
Employing a $k$-nearest neighbours (KNN) classifier in a five-dimensional space (four morphological parameters and redshift), we identify mergers in the mock-observed dataset.
Our analysis reveals a statistically significant enhancement of AGN activity in merging systems, particularly at lower redshifts ($0.5 < z < 0.9$), where central gas reservoirs are more depleted.
This supports the view that mergers contribute more significantly to AGN triggering in environments with low internal gas reservoirs, while their impact may be less pronounced in gas-rich systems.
However, when relying solely on morphological classifications from mock observations, the observed AGN-merger connection weakens, especially at higher redshifts.
This underscores the challenges in detecting merger-induced AGN activity observationally and highlights the importance of combining simulations with realistic mock observations to fully understand the AGN-merger relationship.
\end{abstract}

\begin{keywords}
(galaxies:) quasars: general -- galaxies: interactions -- galaxies: active  
\end{keywords}

\section{Introduction} \label{sec:intro}
Supermassive black holes (SMBHs) are now thought to reside at the centres of nearly all massive galaxies \citep{Kormendy1995,Richstone1998}. 
Though they constitute an insignificant fraction of their host galaxies by volume or mass, their properties appear to be intimately linked with galaxy-scale properties \citep{Kormendy2013}.
In particular, the mass of a SMBH correlates strongly with the stellar mass and velocity dispersion of the host galaxy’s spheroidal component, suggesting a co-evolutionary connection \citep[e.g.][]{Magorrian1998,Ferrarese2000,Gebhardt2000,Marconi2003}.
During episodes of rapid accretion, SMBHs can inject enormous amounts of energy into their surroundings through radiation and outflows, a process collectively known as active galactic nucleus (AGN) feedback \citep{Fabian2012,King2015}.
This energy output can heat or expel the surrounding gas, thereby suppressing further star formation and even halting additional black hole growth \citep{DiMatteo2005,Somerville2015,Fiore2017,Harrison2017,Harrison2018}. 
These feedback mechanisms offer a plausible explanation for how SMBHs can exert significant influence over the evolution of their host galaxies \citep{Booth2009,Dave2019,Nelson2019,Wellons2023}.

Conversely, the evolution of galaxies can also influence the growth of central black holes, particularly through galaxy mergers. 
One of the earliest simulation-based studies to propose such a connection was conducted by \citet{Barnes1991}. 
They demonstrated that during a major merger, strong tidal interactions can induce bar formation, which in turn drives gas into the central regions of galaxies, assembling dense nuclear gas clouds that potentially fuel black hole accretion and star formation.
Since then, mergers have been widely considered one of the primary mechanisms for funneling gas toward central black holes and triggering AGN activity \citep[e.g.][]{Gebhardt2000,Kauffmann2000,Cattaneo2005,Wild2007,Somerville2008}.
Simulations of major mergers between gas-rich disc galaxies have consistently demonstrated that gravitational torques during the interaction can drive substantial gas inflows toward the galactic centre, fueling the central SMBH \citep{Springel2005b, Springel2005a, DiMatteo2005,Cox2006,Hopkins2006,Cox2008}.
The cosmological models of \citet{Hopkins2008} and \citet{Somerville2008} further demonstrated that assuming major mergers are a primary trigger for luminous AGN activity can naturally reproduce the observed evolution of AGN luminosity density and clustering from $z=0$ to 6, supporting mergers as a dominant driver of luminous AGN \citep[see also][]{Hirschmann2012}.
More recently, simulation-based statistical analyses using the IllustrisTNG cosmological hydrodynamic simulations have confirmed that both galaxy pairs and post-merger systems exhibit a significantly enhanced AGN fraction compared to matched control samples \citep{Quai2023, Byrne-Mamahit2023, Byrne-Mamahit2024,Schechter2025}.
These findings provide further support for the merger-driven AGN triggering scenario within the context of cosmological simulations.

Evidence for a connection between AGN activity and galaxy mergers has also emerged from observational studies, beginning with small samples that identified morphological signs of interactions in AGN host galaxies \citep{Surace1998, Canalizo2001, Jahnke2004}. These early works found that many AGNs, particularly luminous ones, reside in galaxies showing tidal features or nearby companions.
Later studies based on larger samples provided more statistically robust support. \citet{Alonso2007} found a modest increase in [OIII]-selected AGNs among close galaxy pairs, especially those with strong tidal features. \citet{Koss2010}, using a hard X-ray-selected AGN sample at $z < 0.05$, showed that AGNs are more common in galaxies with nearby companions or disturbed morphologies. \citet{Ellison2011}, analyzing over 10,000 SDSS galaxies, reported a significantly enhanced AGN fraction in close pairs ($<10 \thinspace \rm kpc$), particularly in equal-mass systems.
Further support comes from \citet{Silverman2011}, who examined 562 massive galaxies in close kinematic pairs from the zCOSMOS survey ($0.25 < z < 1.05$) and found that the AGN fraction is boosted by a factor of $\sim$2 compared to isolated galaxies.
Collectively, these large-sample studies provide statistical support for a link between mergers and enhanced AGN activity, particularly in close galaxy pairs.

However, this AGN-merger connection is not universally observed; several observational studies have found little or no evidence that AGN activity is significantly enhanced in merging or interacting galaxies.
\citet{Schawinski2012} examined heavily obscured quasars at $z\sim 2$ and found that the majority of their host galaxies are disc-dominated rather than visually identifiable merger remnants, suggesting that secular processes play a key role in black hole growth at this epoch.
\citet{Villforth2014} studied the morphologies of X-ray-selected AGN host galaxies at $0.5<z<0.8$ and found no significant excess of merger features compared to control galaxies, concluding that major mergers account for fewer than 6 \% of AGN in their sample.
Similarly, \citet{Rosario2015} showed that while AGNs at $z\sim 1$ exhibit mildly elevated disturbance signatures, AGN hosts at $z\sim 2$ show no clear excess compared to inactive galaxies, suggesting a declining role of mergers in AGN triggering at higher redshift.
\citet{Villforth2017} further focused on the most luminous quasars at $z\sim 0.6$ and found that major mergers are still not the dominant mechanism for triggering black hole accretion.
A similar conclusion was reached by \citet{Mechtley2016}, who analyzed rest-frame $V$-band images of $z\sim 2$ quasars and found only a slight enhancement of merger signatures in the quasar sample, indicating that gas-rich mergers are not the dominant channel for black hole activation at this epoch.
More recently, \citet{Omori2025} analysed the connection between merger-SFR-AGN using the low-$z$ ($0.01<z<0.35$) HSP-SSC galaxies and concluded that secular processes are an important driver of star formation as well as AGN activity.
Indeed, AGN outflow activity is widely observed in relatively normal gas-rich systems at $z\sim 1-2$ \citep{Genzel2014}.

Some of these apparently contridictory results can be understood by realizing that the observed connection between AGNs and galaxy mergers can strongly depend on how AGNs are selected and defined.
Different AGN selection methods, such as optical emission lines, X-ray luminosity, or mid-IR colours, probe different populations due to varying levels of obscuration.
For example, \citet{Koss2010} found a stronger merger association in hard X-ray-selected AGNs than in optically selected ones, suggesting that obscuration can mask AGN signatures in some wavelengths.
\citet{Fan2016} analysed highly obscured, IR-luminous AGNs (“Hot DOGs”) at $z\sim3$ and found a high merger fraction ($\sim 62$\%), in contrast to the lower merger incidence in unobscured, UV/optical-selected AGNs.
These results align with the idea that obscured and highly luminous AGNs show a stronger link to mergers \citep[e.g.,][]{Treister2012}.
\citet{Pfeifle2023} used the XMM-Newton and NuSTAR observation for the mid-IR selected AGNs to infer the column densities.
The densities are as high as $\gtrsim 10^{24}\,\rm cm^{-2}$, again confirming the relationship between late-stage mergers and highly obscured AGNs.
Indeed, several more observations have shown the connectivity between MIR-selected AGNs and galaxy mergers (either the existence of the close-pairs or the post-merger signatures) \citep{Weston2017, Donley2018, Gao2020, Secrest2020, Comerford2024, Bonaventura2025}.

Additional ambiguity arises from the difficulty of reliably identifying galaxy mergers. 
Tidal features, one of the clearest indicators of recent interactions, often have low surface brightness and can easily be missed, especially at high redshift or in shallow imaging.
Projection effects and varying viewing angles further complicate the interpretation of merger signatures.
Some studies, such as \citet{Kocevski2012} and \citet{Rosario2015}, have attempted visual classification to identify disturbed morphologies, but such methods remain subjective and sensitive to data quality.
Taking advantage of publicily released large-scale simulation datasets, recent studies have proposed methods to detect mergers with high precision through mock observations of these galaxies \citep{Pearson2019, Ferreira2020, Guzman-Ortega2023}.

In addition to selection effects, there are physical reasons that mergers and AGNs may not appear connected in observations.
During the peak of morphological disturbance, the central engine may still be deeply buried in dust and gas, rendering the AGN undetectable in optical or even X-ray wavelengths \citep[e.g.,][]{Sanders1988,Kocevski2015,Ellison2019}.
\citet{Dougherty2024} selected AGNs that are observed in infrared but not in X-rays and concluded that only the fraction of these most obscured AGNs have an elevated probability of having a close companion.
Powerful AGN- or starburst-driven outflows may be required to clear the obscuring material, delaying AGN detectability.
Moreover, simulations indicate a time lag between the initial merger stages (when tidal features are strongest) and the peak of black hole accretion, due to the time required for gas to lose angular momentum and reach the nucleus \citep{Hopkins2008,Johansson2009,Choi2014}.
These factors can lead to an observational mismatch between merger features and AGN activity.
Finally, it is clear that not all AGNs are triggered by major/minor mergers: secular processes such as disc instabilities \citep{Bournaud2011} or recycled gas fueling \citep{Ciotti2007,Choi2024} likely also contribute, especially at lower luminosities or at different cosmic epochs.

Although the two approaches -- comparing the AGN (merger) fraction of merger (AGN) and control samples -- are often framed as addressing the same “AGN–merger connection,” they in fact quantify different conditional probabilities.
The AGN-fraction approach compares the probability of observing AGN activity in merging (or interacting) systems relative to non-merging controls, and thus most directly probes whether mergers are associated with an elevated incidence of accretion at fixed host properties.
In contrast, the “merger-fraction” approach measures the fraction of AGN hosts that are classified as mergers or exhibit merger/disturbance signatures, and is therefore better interpreted as the prevalence of merger signatures within an AGN-selected sample rather than the causal  contribution of mergers to triggering.
Because these two quantities are related by Bayes’ theorem and depend on the underlying base rates, a modest merger-associated enhancement in the AGN incidence does not necessarily imply a large excess of merger signatures in AGN-selected samples.
This is particularly relevant if merger-driven fueling is sub-dominant to secular channels and/or if the observability windows for AGN activity and merger signatures are mismatched.
Consistent with this distinction, many observational studies reported no elevated disturbance fractions of AGN-selected samples over inactive control galaxies \citep{Cisternas2011, Kocevski2012, Villforth2014,Mechtley2016,Marian2019} concluding that mergers are not the dominant contributor to AGNs.
We therefore treat both approaches as complementary, but not interchangeable, diagnostics and evaluate them jointly.

In this study, we aim to assess the strength and observability of the AGN–merger connection using high-resolution cosmological zoom-in simulations of massive galaxies at $0.5 < z < 3$.
We first investigate how frequently AGN activity follows merger events in the simulation data, using a statistically consistent framework motivated by observational studies.
We then examine how this intrinsic connection may be altered or suppressed when subject to observational limitations such as morphological ambiguity.
To do so, we generate JWST mock observational images of the same galaxies using radiative transfer calculations that incorporate realistic observational effects.
By comparing the AGN-merger connection identified directly from the simulations with that inferred from the mock images, we evaluate to what extent observational biases may hinder the detection of merger-driven AGN activity.
The general procedure is similar to what has been already done in \citet{Sharma2024}, while the details of the definition of AGNs and galaxy mergers and the mock observation scheme are different.
The results will be directly compared in Section \ref{sec:compare_prev}.

Our paper is outlined as below: in Section \ref{sec:simul}, the basic features of the simulation used in this work are briefly summarised.
In Section \ref{sec:intrinsic}, we introduce how the galaxy mergers and the AGNs are defined from the simulation data, and show
their statistical connectivity.
The whole procedure of conducting the mock observations and the observability of the signal is described in Section \ref{sec:mock_obs}.
After discussing the impact of using different merger definitions in Section \ref{sec:discussion}, we summarize our findings in Section \ref{sec:summary}.


\section{Data} \label{sec:simul}
We employ a set of cosmological zoom-in simulations originally presented in \citet{Choi2017}, which follow the formation and evolution of massive elliptical galaxies with $M_{\star} \gtrsim 10^{11}\,h^{-1}\,\rm M_{\odot}$ at $z=0$.
The high-resolution initial conditions are based on the work of \citet{Oser2010}, constructed by identifying target halos from a dark matter-only simulation in a comoving box of $\sim 72\,h^{-1}\,\rm Mpc$ (WMAP3 cosmology), and tracing back all particles within $2R_{\rm vir}$ for refinement.

The simulations are run with the {\tt SPHGal} code \citep{Hu2014}, a modified version of GADGET-3 \citep{Springel2005c} that incorporates a pressure-entropy SPH formulation, improved artificial viscosity, and thermal conduction to address fluid mixing issues in classical SPH. 
The mass resolution is $m_{\star,\rm gas} = 4.2 \times 10^6 h^{-1} M_{\odot}$ for baryonic particles and $m_{\rm DM} = 2.5 \times 10^7 h^{-1} M_{\odot}$ for dark matter.
The corresponding comoving gravitational softening lengths are $\epsilon_{\star,\rm gas} = 0.4 h^{-1} \,\rm kpc$ and $\epsilon_{\rm DM} = 0.89 h^{-1} \rm kpc$.

Star formation and chemical enrichment follow the prescription in \citet{Aumer2013}, with contributions from Type Ia and Type II supernovae and AGB stars \citep{Iwamoto1999, Woosley1995, Karakas2010}.
Eleven chemical species (H, He, C, N, O, Ne, Mg, Si, S, Ca, Fe) are individually tracked for gas and star particles.
Turbulent diffusion allows metal mixing among gas particles, enhancing the realism of chemical evolution.
Stellar feedback is modeled following \citet{Nunez2017}, incorporating radiation and winds from young stars, multi-phase SN feedback, and AGB winds.

Black holes are seeded with an initial mass of $10^5 h^{-1} \rm M_{\odot}$ at the centre of halos once their virial mass exceeds $10^{11} h^{-1} \rm M_{\odot}$.
Black holes grow via two mechanisms: gas accretion and mergers.
Gas accretion is modeled using the Bondi-Hoyle-Lyttleton formalism \citep{Hoyle1939, Bondi1944, Bondi1952}, with a soft Bondi criterion \citep{Choi2012,Choi2015} that accounts for the geometric overlap between gas particle volumes and the Bondi radius, as well as a free-fall timescale correction.
The accretion rate is not explicitly capped at the Eddington limit; instead, the simulation incorporates the effect of the Eddington force, which exerts a radial outward push on the electrons associated with the gas particles near the black hole.
BH-BH mergers are allowed when the particles are within the smoothing lengths of each other and their relative velocity is below the local sound speed \citep{Springel2005b}.
Because black holes are repositioned to the nearby potential minimum at every timestep, the BH-BH merger occurs right after their hosts merge.

The simulations implement a mechanical AGN feedback model that launches winds near the black hole at a fixed velocity of $v_{\rm outf,AGN} = 10,000$ \kms, with the mass loading set by the inflow rate and a feedback efficiency parameter $\epsilon_{\rm w} = 0.005$ \citep{Choi2017}.
This feedback imparts mass and momentum to the surrounding gas \citep{Choi2012,Choi2014}, effectively quenching star formation and self-regulating black hole growth \citep{Ostriker2010,Choi2015}.
The model reproduces observed high-velocity outflows \citep[e.g.][]{Arav2020} expected from radiatively efficient accretion \citep[e.g.][]{Proga2000} and employs a time-step limiter to ensure accurate shock propagation.
In addition, Compton and photoionization heating, along with the associated radiation pressure from moderately hard X-rays ($\sim$50 keV) \citep{Sazonov2004,Sazonov2005}, are included.
Combined, these mechanical and radiative AGN feedback processes as well as stellar feedback efficiently suppress star formation in massive galaxies, consistent with previous simulation results \citep{Choi2015,Choi2017,Kim2025}.

We analyze simulation snapshots within the redshift range $0.5 < z < 3$, with 42 snapshots available for each target galaxy.
This results in a total of 1302 galaxy–snapshot data points, which we divide into three redshift bins for analysis: 434 in the low redshift bin ($0.5 < z < 0.9$), 403 in the mid-redshift bin ($0.9 < z < 1.5$) and 465 in the high-redshift bin ($1.5 < z < 3.0$).
The median values of the stellar mass at each redshift bin corresponds to $4.43$, $8.04$ and $9.44\times10^{11} \,M_{\odot}$ from high to low-z bins, though higher redshift bins gradually have more galaxies in low-mass tail.

While the snapshots are saved at relatively coarse time intervals, the simulations additionally record the detailed growth history of black holes at much finer timesteps during run-time.
These high-resolution black hole data are used to define AGN activity, as described in Section~\ref{sec:intrinsic_agn_definition}.


\section{Intrinsic AGN-Merger Connection in Simulations} \label{sec:intrinsic}
To assess whether galaxy mergers play a causal role in triggering AGN activity, we first utilize the full three-dimensional information available from the simulations.
This section describes how mergers and AGNs are defined using intrinsic galaxy properties, and presents a statistical analysis of their temporal connection.
The methodology follows the commonly adopted framework in observational studies, allowing a direct comparison with the results derived from mock observational data in the next section.

\subsection{Merger Definition}\label{sec:intrinsic_merger_definition}
To investigate whether galaxy mergers intrinsically trigger active galactic nuclei (AGNs), we first define mergers using the full six-dimensional phase-space information available from the simulations.
We construct individual merger trees for each of the 31 simulated galaxies using the {\tt Rockstar} halo finder \citep{Behroozi2013} in combination with the {\tt consistent-trees} algorithm \citep{Behroozi2012}, which allows robust tracking of halo coalescence events.
At each merger node in the tree, we compute the stellar mass ratio between the merging galaxies at one snapshot before the event, corresponding to $\sim 140\,\rm Myr$ on average.
Merger events with a stellar mass ratio greater than or equal to 0.25 are classified as major mergers, while those with a ratio greather than or equal to 0.1 are classified as major+minor mergers.

To account for potential time delays between merger events and the onset of AGN activity, we introduce a temporal window around each merger.
In the post-merger-only case, we consider a time interval $[0, \Delta t_{\rm merger}]$ immediately following coalescence.
To complement this post-merger-only definition and incorporate possible contributions from pre-merger phases, we also consider a symmetric time window around coalescence, $[-\Delta t_{\rm merger}, \Delta t_{\rm merger}]$.
These time windows allow us to probe the temporal association between mergers and AGN activity across different evolutionary stages.
A galaxy at a given snapshot is classified as being in a merger phase only if it satisfies both the stellar mass ratio and the time-window criteria described above; otherwise it is defined as a non-merger.
In particular, interactions with mass ratios below our major-merger threshold (i.e. minor mergers) are included in the non-merger category.

\begin{figure*}
    \centering    \includegraphics[width=\linewidth]{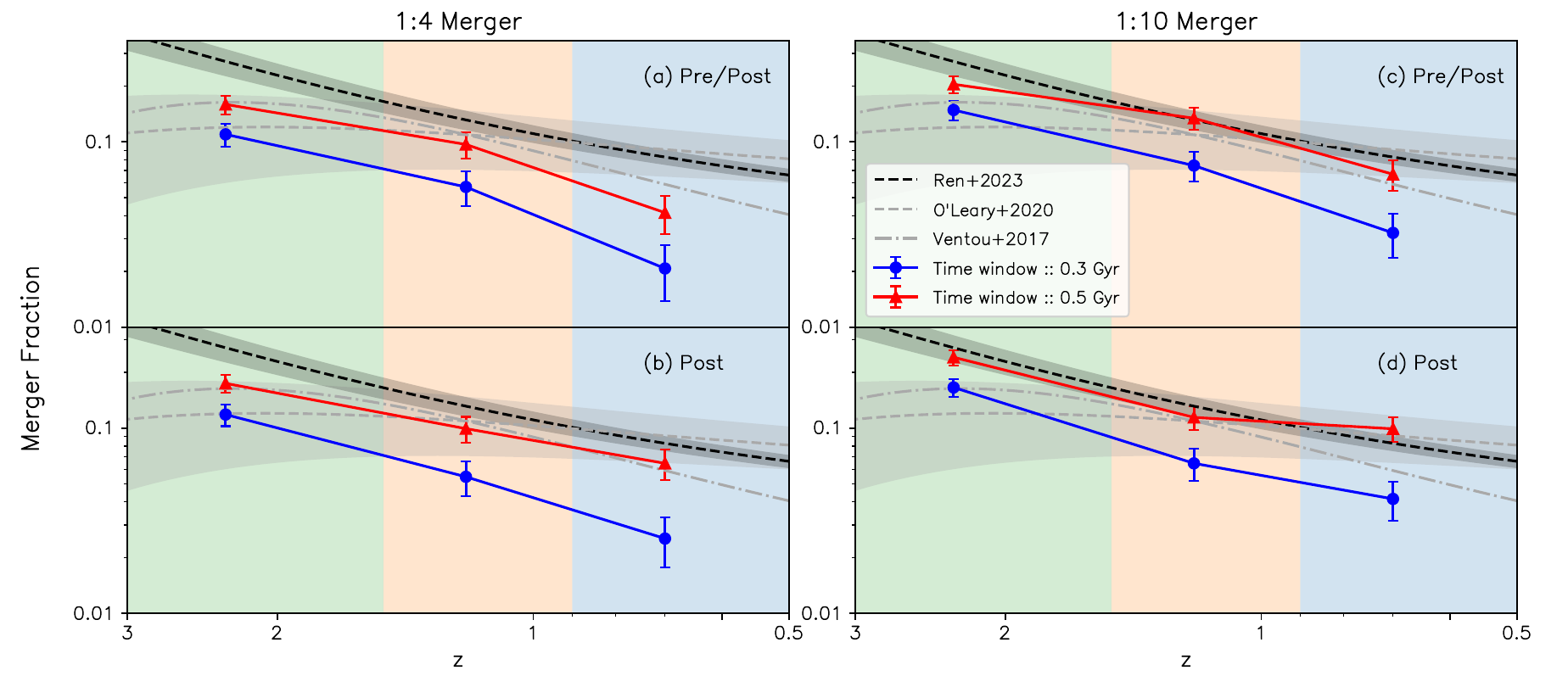}
    \caption{
    Redshift evolution of galaxy merger fractions under different merger definitions. 
    The error bars denote the 68\% Bayesian credible intervals computed from the quantiles of the beta distribution \citep{Cameron2011}.
    We test eight definitions that vary in (i) stellar mass ratio thresholds (major vs. major+minor; \textit{left} vs. \textit{right}), (ii) time window symmetry (post-merger only vs. pre+post-merger; \textit{upper} vs. \textit{lower}), and (iii) time window width (0.3 Gyr vs. 0.5 Gyr; \textit{blue} vs. \textit{red}).
    Observational results include visual classifications from \citet{Ren2023} (black dashed lines) and close-pair statistics from \citet{Ventou2017} (gray dot-dashed). \citet{OLeary2021} (gray dashed) shows merger fractions derived from an empirical model.
    }
    \label{fig:merger_frac_z_func}
\end{figure*}

Figure~\ref{fig:merger_frac_z_func} shows the merger fraction as a function of redshift, computed using a range of merger definitions based on stellar mass ratio thresholds, time window widths, and whether to include pre-merger phases.
We compare these simulation-based estimates to observational measurements.
Because our merger definitions involve the post-merger phases, we assume that the visual classification of \citet{Ren2023} (the black dashed line) may be more relevant to our results than the close-pair statistics of \citet{Ventou2017} (the gray dot-dashed line).
Given that our simulation sample targets a specific subset of massive galaxies and that observational merger classifications rely on limited and often projected information, a one-to-one correspondence between the two is not anticipated. Nevertheless, several of our merger definitions yield merger fractions in reasonable agreement with observed trends.

Unless stated otherwise, we adopt as our fiducial merger definition the one marked in red in panel (b): major mergers within a post-merger time window of $\Delta t_{\rm merger} = 0.5 \rm Gyr$, following various works such as \citet{Lotz2010}.
According to \citet{Ellison2025}, while the AGN excess is maximized immediately after the coalescence ($\Delta t_{\rm merger}<0.16\,\rm Gyr$), there is still an expected signal when using the $\Delta t_{\rm merger}\sim 1\,\rm Gyr$.
We examine the sensitivity of the AGN-merger connection to alternative merger definitions in Section~\ref{sec:discussion}.

\subsection{AGN Definition} \label{sec:intrinsic_agn_definition}

\begin{figure}
 \centering
 \includegraphics[width=\linewidth]{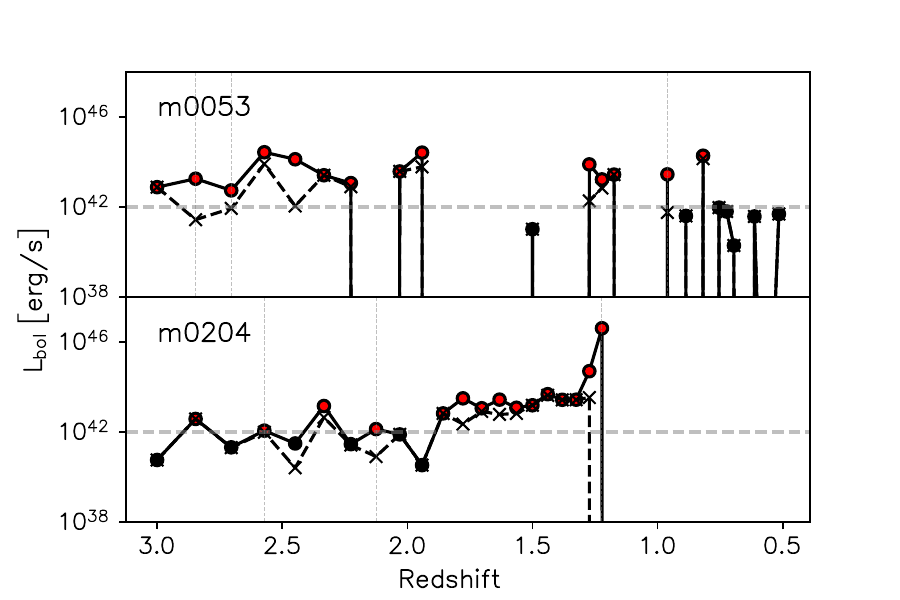}
 \caption{Redshift evolution of the bolometric luminosities of central SMBHs for two example galaxies.
The instantaneous luminosity measured at each simulation snapshot is shown as dashed lines, while the maximum luminosity within the preceding $10^5\,\rm yr$ is shown as solid lines.
Snapshots classified as AGN based on whether the luminosity exceeded the threshold at any point within the past $10^5\,\rm yr$ are marked with red circles.
Gray vertical dashed lines indicate cases where AGN activity would be missed if only the instantaneous luminosity were considered, but is recovered using the past-activity-based definition.}
 \label{fig:Lbols}
\end{figure}

When gas is accreted onto a BH at the Bondi-Hoyle-Lyttleton accretion rate ($\dot{M}_{\rm BHL}$), a fraction $\eta$ of the rest mass energy is converted into radiation and BH mass grows at $\dot{M}_{\rm BH}=(1-\eta)\dot{M}_{\rm BHL}$.
Assuming a canonical radiative efficiency of $\eta=0.1$, the bolometric luminosity is given by
\begin{equation}
 \label{eq:Lbol}
 L_{\rm bol} = \frac{\eta}{1-\eta} \dot{M}_{\rm BH}c^2
\end{equation}
where $\dot{M}_{\rm BH}=(1-\eta)\dot{M}_{\rm BHL}$.
We classify a galaxy as hosting an active nucleus if its SMBH reaches a bolometric luminosity above a threshold $L_{\rm thr}$.
Following \citet{Singh2023}, we adopt $L_{\rm thr}=10^{42}\,\rm erg\,s^{-1}$ as our fiducial threshold. We verify that our results (Section~\ref{sec:intrinsic_stat}) remain qualitatively unchanged when adopting alternative criteria, such as $L_{\rm thr} = 10^{43}\,\rm erg\,s^{-1}$ or an Eddington ratio threshold of $\lambda_{\rm Edd} \geq 10^{-3}$.

This definition corresponds to the phase when the black hole is radiatively efficient, typically emitting in UV and X-ray bands.
However, AGN activity may also be inferred from narrow-line region (NLR) emission, excited by ionizing radiation from past accretion episodes.
In this case, the SMBH can be identified using diagnostics such as those described in \citet[][BPT diagram]{Baldwin1981}.
\citet{Schawinski2015} estimate the characteristic AGN lifetime relevant to NLR diagnostics as $\Delta t_{\rm AGN} \sim 10^5\rm yr$, based on the typical NLR size ($\sim 10^4$ light-years) and the observed fraction of optically elusive AGNs in the Swift BAT sample ($\sim$5\%).

Leveraging the high time-resolution black hole accretion data recorded during simulation runtime (see Section~\ref{sec:simul}), we define a SMBH as \emph{active} at a given snapshot if its bolometric luminosity has exceeded $L_{\rm thr}$ at least once within the preceding $\Delta t_{\rm AGN} = 10^5\rm yr$, regardless of its instantaneous accretion rate.

Figure~\ref{fig:Lbols} illustrates the bolometric luminosity histories of two example galaxies.
Dashed lines represent the instantaneous luminosities at each snapshot, while solid lines trace the maximum luminosity within the preceding $\Delta t_{\rm AGN} = 10^5\,\rm yr$.
Red circles mark the snapshots identified as AGN based on whether the luminosity exceeded the threshold at any point within this lookback window. 
This figure summarizes our AGN definition based on accretion history over a $10^5 \rm yr$ timescale, highlighting the importance of incorporating recent accretion activity beyond the instantaneous state.


\subsection{Intrinsic AGN-Merger Connection}\label{sec:intrinsic_stat}
Using the intrinsic fiducial definition of galaxy mergers (post major merger of $500\,\rm Myr$ time window) and AGN described in subsections \ref{sec:intrinsic_merger_definition} and \ref{sec:intrinsic_agn_definition}, we analysed the statistical connection between mergers and AGNs based on simulation data.
We note that four data points in the highest redshift bin have been excluded from this analysis because their mophological parameters cannot be calculated due to their low surface brightness (see Section ~\ref{subsec:morph}).
Following methodologies commonly adopted in observational studies, we perform a symmetric comparison in two directions:
(1) we measure the AGN fraction among merging and non-merging galaxies, and 
(2) we compare the merger fraction among AGNs and non-AGNs.
A key requirement in this type of analysis is the use of well-matched control samples.
To ensure a fair comparison, we construct redshift- and stellar-mass-matched control sets for each data point.
More specifically, if a galaxy with a given stellar mass in a given redshift bin is identified as a merger, we randomly select corresponding control samples that are classified as non-mergers, lie in the same redshift bin, and have stellar masses within $0.1\,\mathrm{dex}$.
In this process, we do not restrict the control galaxies to come from the same zoom-in box, but instead allow candidates from all of our zoom-in simulations.
For AGN activity, we adopt an analogous procedure, matching each AGN snapshot to non-AGN control snapshots in the same redshift bin and within $0.1\,\mathrm{dex}$ in stellar mass, again allowing candidates from all zoom-in runs.
To account for the statistical uncertainty inherent in the control sample selection, we repeat the procedure for 1000 random realizations and incorporate the resulting $1\sigma$ errors into our total error budget.

\begin{figure}
    \centering
    \includegraphics[width=\linewidth]{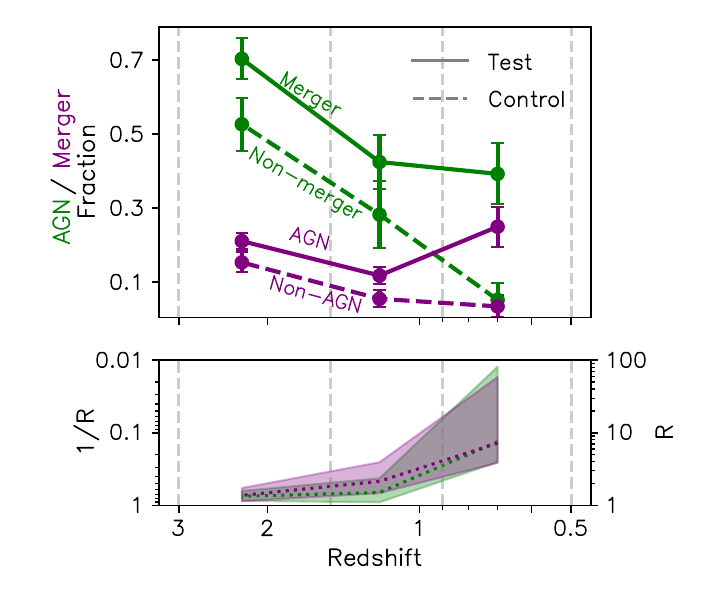}
    \caption{
    Intrinsic connection between galaxy mergers and AGN activity across redshift. The upper panel shows the AGN fraction among merging galaxies (solid green) compared to their non-merging control counterparts (dashed green), and the merger fraction among AGNs (solid purple) compared to non-AGN controls (dashed purple).
    The lower panel presents the inverse enhancement ratio ($1/R$) for both cases, along with corresponding error bars.
    Although most observational studies present the enhancement in terms of $R$, we use $1/R$ to avoid divergence issues when the control sample fractions are small.
    For convenience, the equivalent $R$ values are also indicated on the right-hand $y$-axis, inverted for clarity.}
    \label{fig:stat_intrinsic}
\end{figure}

\begin{table}
\centering
\caption{{\bf Quantitative Summary of the Intrinsic AGN–Merger Connection} : This table lists the numerical values shown in Figure~\ref{fig:stat_intrinsic}, including AGN and merger fractions for both test and control samples across redshift bins.}
\resizebox{\linewidth}{!}{%
\begin{tabular}{lc|ccc}
\hline
                                                       & \multicolumn{1}{l|}{} & High-$z$      & Mid-$z$       & Low-$z$       \\
                                                       & \multicolumn{1}{l|}{} & ($3<z<1.5$)   & ($1.5<z<0.9$) & ($0.9<z<0.5$) \\ \hline
\multicolumn{1}{l|}{\multirow{2}{*}{$f_{\rm AGN}$}}    & merger                & 0.70$\pm$0.06 & 0.42$\pm$0.07 & 0.39$\pm$0.08 \\
\multicolumn{1}{l|}{}                                  & non-merger            & 0.52$\pm$0.07 & 0.28$\pm$0.09 & 0.05$\pm$0.05 \\ \hline
\multicolumn{1}{l|}{\multirow{2}{*}{$f_{\rm merger}$}} & AGN                   & 0.21$\pm$0.02 & 0.12$\pm$0.02 & 0.25$\pm$0.05 \\
\multicolumn{1}{l|}{}                                  & no-AGN                & 0.15$\pm$0.03 & 0.05$\pm$0.02 & 0.03$\pm$0.03 \\ \hline
\end{tabular}%
}
\label{tab:stat_intrinsic}
\end{table}

Figure~\ref{fig:stat_intrinsic} presents the intrinsic connection between mergers and AGN activity across redshift.
Solid lines represent the test samples (mergers or AGNs), while dashed lines correspond to their redshift-$M_\star$-matched control samples.
We distinguish the two directions of comparison by colour: green lines show AGN fractions among merging and non-merging galaxies, while purple lines represent merger fractions among AGNs and non-AGNs.

Since the total number of mergers is not the same after the mock observation (as discussed later in Section~\ref{sec:synthetic_stat}), we additionally present in the lower panel the ratio between control and test sample fractions, defined as $1/R = y_{\rm ctr} / y_{\rm test}$.
The associated uncertainty is computed via error propagation:
\begin{equation}
    \sigma_{1/R} = \frac{1}{R}\sqrt{\left(\frac{\sigma_{y_{\rm ctr}}}{y_{\rm ctr}}\right)^2 + \left(\frac{\sigma_{y_{\rm test}}}{y_{\rm test}}\right)^2}.
\end{equation}
While previous studies typically report enhancement as $R = y_{\rm test} / y_{\rm ctr}$, this can lead to large and unstable error bars when $y_{\rm ctr}$ is small, as seen in our results.
We therefore plot $1/R$ for numerical stability, and include the corresponding $R$ values on the right-hand y-axis, which is inverted for clarity.
Note that the error bars correspond only to the $1/R$ axis and should not be interpreted on the $R$ scale.

Figure~\ref{fig:stat_intrinsic} shows that galaxies undergoing mergers are more likely to host AGNs across all redshift bins.
At high redshift ($1.5 < z < 3$), over 70 \% of merging galaxies host AGNs, while the non-merging control sample also shows a relatively high AGN fraction (50 \%), resulting in only a modest enhancement ($R \lesssim 1.5$).
This suggests that mergers are not the dominant driver of SMBH fueling at this epoch.

In contrast, at lower redshift ($0.5 < z < 0.9$), although the overall AGN fraction declines, the enhancement among mergers becomes much more significant, reaching $R \sim 10$.
A similar trend is seen in the merger fraction among AGN vs. non-AGN samples: while mergers are more likely to be found in AGNs at all redshifts, the excess is strongest at low redshift, again with $R \sim 10$.

\begin{figure}
    \centering
    \includegraphics[width=0.85\linewidth]{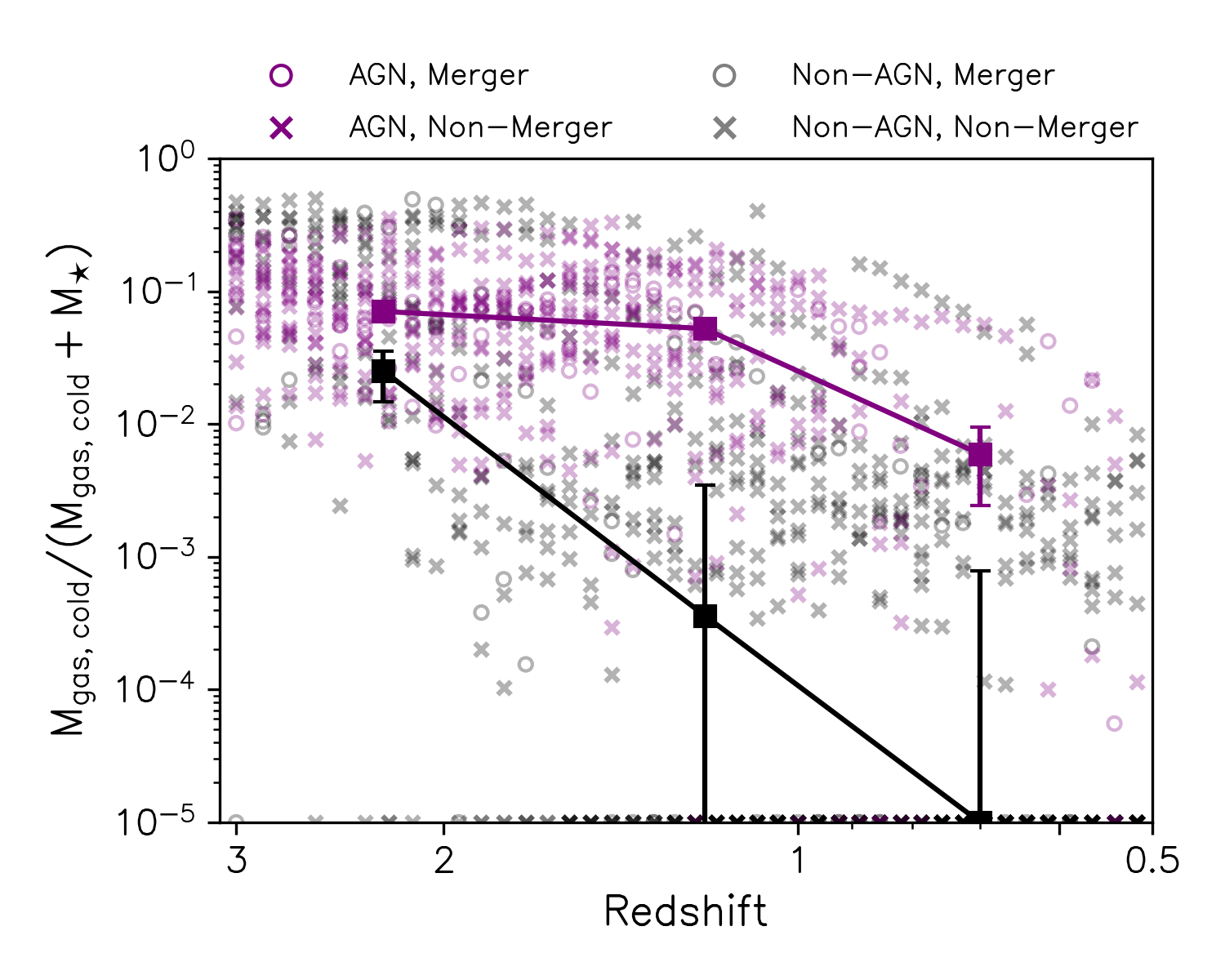}
    \caption{
    Cold ($<10^4\,\rm K$) gas fractions (within $<0.1\,R_{\rm vir}$) relative to the combined mass of cold gas and stars for all galaxies in our sample. Purple symbols denote AGN hosts and grey symbols non-AGN galaxies. Mergers and non-mergers are distinguished by circles and crosses, respectively. The solid line represents the median value at each redshift bin.}
    \label{fig:gas_fraction}
\end{figure}

We interpret the redshift-dependent strength of the AGN–merger connection as a consequence of the evolving central gas reservoir in galaxies.
We define the central gas fraction as the ratio of gas mass to the combined mass of cold gas ($<10^4\,\rm K$) and stars within the inner $0.1\,R_{\rm vir}$ of each galaxy:
\begin{equation}
    f=\frac{M_{\rm gas,cold}}{M_{\rm gas,cold}+M_{\star}}
\end{equation}

Figure~\ref{fig:gas_fraction} shows the redshift evolution of central cold gas fractions for our galaxy sample, where AGNs are marked as purple and non-AGNs as black.
Comparing the median of the cold gas fraction (solid lines) of each group, AGNs show relatively higher gas fractions.

Figure~\ref{fig:gas_fraction} shows the redshift evolution of central cold gas fractions for our galaxy sample.
Galaxies display a steady decline in cold gas fraction toward lower redshift, consistent with the broader cosmological trend of gas depletion over time.
At higher redshifts, galaxies are generally more gas-rich, providing abundant fuel for black hole accretion even in the absence of mergers.
Notably, AGN hosts (purple symbols) tend to be among the most gas-rich systems at all redshifts.
At high redshift, gas-rich galaxies are more likely to exhibit AGN activity, while in the low-$z$ sample, AGNs are preferentially found in the relatively gas-rich tail of the population.

This evolution in gas content has direct implications for the AGN–merger connection.
In gas-rich systems at high redshift, central SMBHs can be fueled efficiently through internal processes such as disc instabilities or secular inflows, thereby reducing the relative importance of mergers as a triggering mechanism.
Conversely, in gas-poor galaxies at lower redshift, black hole accretion becomes increasingly dependent on external dynamical events, such as mergers, that can deliver fresh gas to the nuclear regions.

In addition to the effect of the overall cold gas reservoir, the efficiency of angular momentum removal due to mergers also depends on the gas fraction.
Idealized simulations and analytic models suggest that the loss of angular momentum during a merger is dominated by internal gravitational torques exerted by the stellar bar on the gas component \citep{Hopkins2012}.
When the gas fraction is very high in a galaxy, the stellar bar weakens or forms later, reducing the merger-driven inflows.
\citet{Schechter2025} also discussed this effect, and showed larger sSFR ehnahcements at lower gas fractions using the TNG50 simulations.


\section{AGN-Merger Connection in Mock Observations}\label{sec:mock_obs}
In the previous section, we demonstrated that galaxy mergers are statistically linked to enhanced AGN activity in our simulations, with the strongest signal observed in the redshift range $0.5 < z < 0.9$.
Given the mixed results reported in observational studies (see Section~\ref{sec:intro}), it is crucial to assess whether this intrinsic connection is detectable under realistic observational conditions.
In this section, we describe how we generate mock JWST-like images from the simulations using radiative transfer techniques, and how galaxy mergers are identified based on image-based morphology.
We then examine how the AGN–merger connection appears when using this observationally motivated merger classification.


\subsection{JWST Mock Image Generation} \label{subsec:RT}
We first generate mock JWST-like images of simulated galaxies using the dust radiative transfer code {\tt powderday} \citep{Narayanan2021}, which combines the stellar population model {\tt FSPS} \citep{Conroy2010} and the radiative transfer code {\tt Hyperion} \citep{Robitaille2011} through a {\tt yt} \citep{Turk2011} frontend.
For each pixel, we compute spectral energy distributions (SEDs) that include contributions not only from the stellar population but also from the AGN.
The AGN UV-optical continuum emission is described by the piecewise power-law distribution model from \citet{Rowan-Robinson1995} and the re-emission from dusty torus is modelled using \citet{Nenkova2008a,Nenkova2008b}.
We refer readers to \citet{Narayanan2021} for implementation details.
These pixel-by-pixel SEDs are then convolved with the JWST/NIRCam F277W filter throughput curve, and the images are rendered at an angular resolution of $0.03''$ on a $512 \times 512$ pixel grid.

\begin{figure*}
    \centering
    \includegraphics[width=0.8\linewidth]{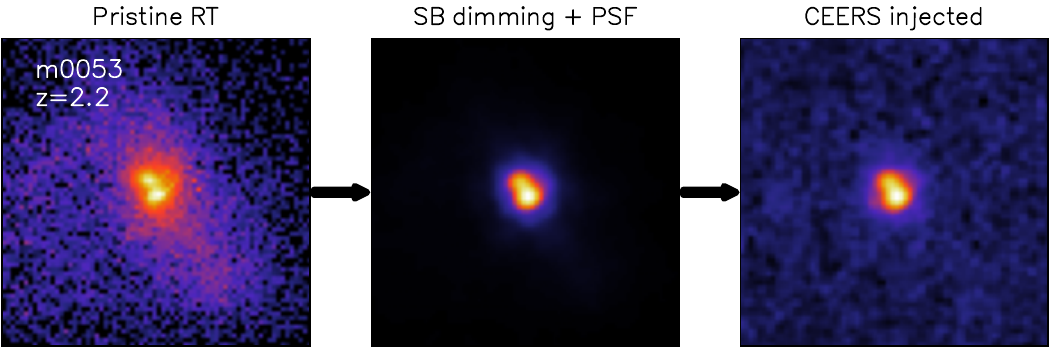}
    \caption{Workflow for generating realistic mock JWST galaxy images.
We first create pristine radiative transfer images using {\tt powderday} (\textit{left}).
Surface brightness dimming is applied according to the cosmological scaling $(1+z)^{-4}$, followed by convolution with the JWST/NIRCam F277W point spread function (\textit{middle}).
A randomly selected $512 \times 512$ patch from the CEERS survey is injected as a background to produce the final mock observation (\textit{right}).
After SB dimming, the brightest pixel values correspond to $\sim 10\,\rm nJy$.}
    \label{fig:post_processing_img}
\end{figure*}

We applied cosmological surface brightness dimming following the standard $(1+z)^{-4}$ law, and convolved each image with the point spread function (PSF) of the F277W filter, simulated using {\tt WebbPSF} \citep{Perrin2015}.
To mimic realistic observational conditions, we utilize fully reduced NIRCam F277W field images from six pointings in the CEERS data release 0.6 \citep{Bagley2023}.
For each mock image, we randomly select a $512 \times 512$ pixel patch from the CEERS background, ensuring that no bright source (above $3\sigma$) is present within the central $256 \times 256$ region.
This patch is added to the radiative transfer image to produce a realistic mock observation, as illustrated in Figure~\ref{fig:post_processing_img}.
For each galaxy at a given snapshot, we generate three mock images from randomly selected orientations, resulting in a total of 3,906 mock images.

\subsection{Morphology-Based Merger Identification} \label{subsec:morph}
To identify mergers from the mock images, we compute four non-parametric morphological parameters using {\tt statmorph} \citep{Rodriguez-Gomez2019}: the Gini coefficient ($G$), the second-order moment of the brightest 20\% of the light ($M_{20}$), asymmetry ($A$), and concentration ($C$).

The Gini coefficient quantifies the inequality of the flux distribution across image pixels. It is defined as the normalized area between the cumulative distribution function of the flux values and that of a uniform distribution.
A value of $G = 0$ indicates perfectly uniform brightness, while $G = 1$ corresponds to maximal concentration in a single pixel. It is computed as:
\begin{equation}
    G=\frac{1}{|\bar{X}|n(n-1)}\sum_{i=1}^n (2i-n-1)|X_i|,
\end{equation}
where $X_i$ are the pixel flux values sorted in increasing order, and $|\bar{X}| = \sum_i |X_i| / n$.

The $M_{20}$ parameter measures the normalized second-order moment of the brightest 20\% of the galaxy's flux.
For a galaxy centered at $(x_c, y_c)$ with $N$ pixels, the total second-order moment is:
\begin{equation}
    \begin{aligned}
    M_{\rm tot} &= \sum_i^N M_{i}=\sum_i^N I_i\left[(x_i-x_c)^2 + (y_i-y_c)^2\right],
    \end{aligned}
\end{equation}
and $M_{20}$ is defined as:
\begin{equation}
    M_{20}=\log \frac{\sum_i M_i}{M_{\rm tot}} {\rm ,\,\,for\,\,}\sum_{i}F_i<0.2 F_{\rm tot},
\end{equation}
where $F_i$ and $F_{\rm tot}$ denote the individual and total fluxes, respectively.

The asymmetry parameter $A$ quantifies rotational symmetry by comparing the galaxy image to its 180$^\circ$ rotated version:
\begin{equation}
    A=\frac{\sum|I_{ij}-I_{ij}^{180}|}{\sum |I_{ij}|}-A_{\rm bkg},
\end{equation}
where $A_{\rm bkg}$ is the background asymmetry measured from blank sky regions.

The concentration index $C$ measures the degree to which light is concentrated in the central region:
\begin{equation}
    C=5\log_{10}\left(\frac{r_{80}}{r_{20}}\right),
\end{equation}
where $r_{80}$ and $r_{20}$ are the radii enclosing 80\% and 20\% of the galaxy's total flux, respectively.
The total flux is measured within 1.5 times the Petrosian radius, and the apertures are centered on the location that minimizes $A$.

In order to calculate the above 4 parameters, we first define the region of interest by detecting and deblending sources from a smoothed image using {\tt photutils} \citep{Bradley2023}.
A Gaussian kernel with full-width at half maximum (FWHM) four times larger than the PSF of the F277W filter is applied to better identify diffuse low-surface-brightness features and suppress noise in the source boundaries.
We successfully compute morphological parameters for 3,894 out of 3,906 mock images.
The remaining 12 images, primarily from the highest redshift bin, have insufficient surface brightness for reliable analysis.
Since these excluded cases do not belong to either the AGN or merger test samples, and control samples outnumber test samples significantly, their omission is not expected to affect the statistical results.

\begin{figure*}
    \centering
    \includegraphics[width=0.9\linewidth]{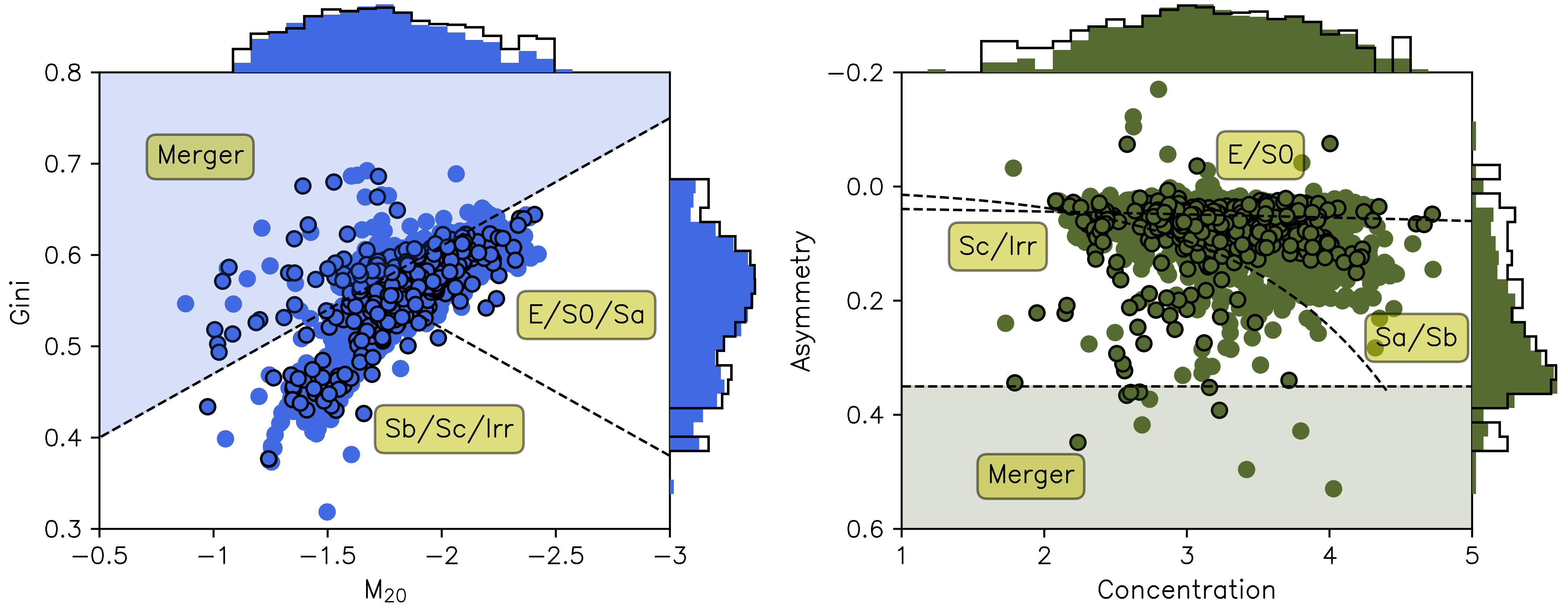}
    \caption{Two-dimensional distribution of four morphological parameters. The dashed lines represent empirical divisions commonly used to distinguish galaxy mergers. Those in the left panel are from \citet{Lotz2008}, while in the right panel they are from \citet{Bershady2000} and \citet{Conselice2003}. Galaxies that are intrinsically classified as mergers are marked with black edges. The histogram for each of the parameters is also shown at the edges. Here, merging galaxies are also identified as black edges. Comparing the dashed lines and the locations of the merger data points, it is clear that the traditional criteria are not providing robust classification of mergers.
    }
    \label{fig:morph_params_2d_hist}
\end{figure*}
Figure~\ref{fig:morph_params_2d_hist} shows the histograms and two-dimensional distributions of the four morphological parameters, with galaxies intrinsically classified as mergers highlighted as black open circles.
It is evident that the commonly used empirical criteria, for example, those proposed by \citet{Lotz2008} and \citet{Bershady2000}, are not effective at recovering mergers in our dataset.
Using either set of criteria, the true positive rate (i.e., the fraction of intrinsic mergers correctly identified) is only $\sim$20\%, while the true negative rate (i.e., non-mergers correctly classified) is approximately 90\%.
This imbalance indicates that traditional morphological thresholds are insufficient for reliable merger identification in our mock observations, as already pointed out in \citet{Abruzzo2018}.

In this work, we first experimented with supervised classifiers that take four morphological summary statistics and redshift as input features.
In particular, we trained random-forest classifiers on $\sim 4{,}000$ mock-observed images, tuning hyperparameters with \texttt{GridSearchCV}.
We found that the cross-validated performance estimates were sensitive to the random seed that determines the CV fold assignments, indicating substantial variance under resampling in this limited-data regime.
To obtain a more stable classifier while staying within the space of interpretable features, we adopt a $k$-nearest neighbours (\texttt{KNN}) algorithm in the five-dimensional space defined by the four morphological parameters and redshift.
Because each prediction is determined by the labels of nearby samples in this feature space, the resulting merger classifier is naturally data-driven and adapts to the local distribution of our mock-observed galaxy sample.


\subsection{Merger Definitions from Multi-dimensional KNN}
\label{subsec:knn}

We classify mergers in a five-dimensional feature space composed of four non-parametric morphological parameters ($G$, $M_{20}$, $A$, and $C$). 
Redshift is included as an input feature to compensate for the use of a single filter (NIRCam F277W) in mock image generation.
As noted in Section~\ref{subsec:RT}, we generated three mock images for each galaxy from different random viewing angles.
To construct training and test sets, we apply an approximate 2:1 split of the total mock images.

A galaxy is classified as a merger if the majority of its $k$ nearest neighbours in this feature space are labelled as intrinsic mergers.
The distance between two points is computed using the Minkowski metric, defined as $[\sum (x_i - y_i)^p]^{1/p}$.
We adopt $p = 1$ (Manhattan distance), which yielded better classification performance in our tests compared to the standard Euclidean distance ($p = 2$).

An important aspect of our dataset is the strong class imbalance: only approximately 10\% of the samples are mergers.
As with most machine learning classification algorithms, this imbalance can lead to biased predictions in the {\tt KNN} classifier, particularly near class boundaries where the majority class (non-mergers) tends to dominate the neighbour votes.

To mitigate this issue, we apply random oversampling of the minority class (mergers) with replacement using the {\tt RandomOverSampler} technique.
We also experimented with the {\tt SMOTE} algorithm, which synthetically generates new samples by interpolating between randomly selected pairs of minority-class points in the feature space.
However, the results were not significantly different from those obtained with simple random oversampling.

The classification procedure involves two key control parameters: {\tt sampling\_strategy}, which specifies the ratio of oversampled minority (merger) samples to majority (non-merger) samples, and $k$, the number of nearest neighbours considered during voting.
To determine the optimal combination of these parameters, we evaluate classification performance across a range of values using multiple metrics.
Since overall accuracy is known to be a poor indicator for imbalanced datasets \citep{deDiego2022}, we instead compare a variety of alternative performance metrics.
Based on this comparison, we select {\tt sampling\_strategy} $= 1.0$ and $k = 11$ as the optimal configuration.
Further details of this parameter selection process are provided in Appendix~\ref{sec:_optimal_params}.

\begin{figure}
    \centering
    \includegraphics[width=\linewidth]{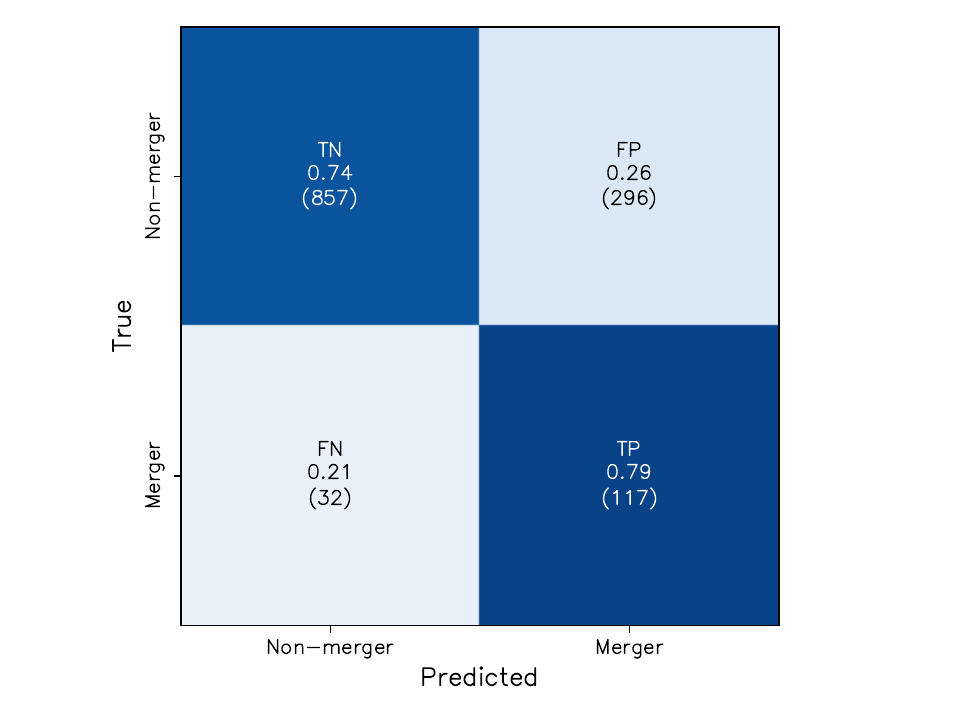}
    \caption{
    Confusion matrix for the test set using the selected {\tt KNN} model ($k=11$, {\tt sampling\_strategy} = 1.0).
    Each entry shows both the number of classified samples and the fraction relative to the true class (row-normalised).}
    \label{fig:confusion_matrix}
\end{figure}

The resulting confusion matrix for our KNN classification is shown in Figure~\ref{fig:confusion_matrix}.
Each cell displays both the number of samples and the corresponding fraction normalised by the total number of samples in each true class (i.e., row-normalised).
The true positive rate (TPR) and true negative rate (TNR), that is, the fractions of mergers and non-mergers correctly classified, reach moderate to high values of approximately 70–80\%.
However, due to the strong class imbalance, the number of false positives (non-mergers misclassified as mergers) remains significantly higher than the number of true positives.
This contamination can affect the analysis of AGN-merger statistics in the next section, potentially diluting the observed connection between mergers and AGN activity.
Compared to the morphology-based random forest classification of simulated post-mergers by \citet{Wilkinson2024}, which reported a completeness ($=\rm TP/(FN+TP)$) of 0.83 and a false positive rate ($\rm FPR=FP/(TN+FP)$) of 0.19 for mock images with a surface-brightness limit of $\sim 27\,\rm mag\,arcsec^{-2}$ and $\rm FWHM\sim 0.25''$, our classifier tends to perform slightly less well primarily due to the strong class imbalance in our dataset.
In particular, their analysis is conducted on a balanced merger/non-merger sample, whereas our {\tt KNN} model is applied in a strongly imbalanced regime.
Though we randomly oversample the merger class, this effectively replicates the same minority samples.
In such a regime, the absolute number of false positives can easily exceed that of true positives, even when the overall completeness and false positive rates are comparable to those obtained in a more balanced dataset.
In addition, differences between our {\tt KNN} classifier and the random-forest model may also contribute to the quantitative discrepancies, since ensemble tree methods can capture more complex, non-linear decision boundaries in the morphology-parameter space.
In this sense, our results indicate that the {\tt KNN} classifier achieves reasonably good performance in terms of completeness and FPR, given the strong class imbalance.
More broadly, as emphasised by \citet{Wilkinson2024}, merger identification based on non-parametric morphology statistics is intrinsically limited.
Because merger observability depends sensitively on viewing angle, a fraction of mergers will remain indistinguishable from non-mergers, setting a fundamental limit on achievable performance.
More broadly, as emphasised by \citet{Wilkinson2024}, merger identification based on non-parametric morphology statistics is expected to have an intrinsic upper limit because the observability of merger features depends sensitively on viewing angles;
our {\tt KNN}-based scheme should therefore be interpreted within this fundamentally limited regime.
Even with these limitations, our classification method
allows us to quantify the level of uncertainty introduced by using incomplete information to identify mergers within the same dataset.

\subsection{Synthetic Statistics and Comparison to Observations}
\label{sec:synthetic_stat}

\begin{figure}
    \centering
    \includegraphics[width=\linewidth]{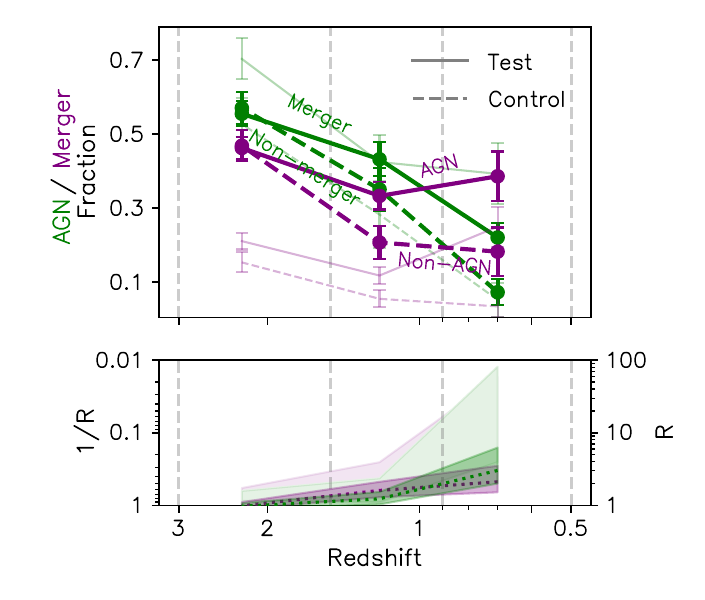}
    \caption{The same as in Figure \ref{fig:stat_intrinsic}, except that the galaxy mergers now indicate {\tt KNN}-identified mergers based on the morphological statistics measured from the mock images. Because {\tt KNN} predicts more mergers, the merger fractions (purple lines) significantly increased compared to the intrinsic cases. The results from Figure \ref{fig:stat_intrinsic} (intrinsic statistics) are overlaid with lighter colored lines and shades.}
    \label{fig:stat_synthetic}
\end{figure}

In the previous section, we classified galaxy mergers using a KNN model based on redshift and four non-parametric morphological parameters.
We now reapply the statistical analysis of AGN-merger connection from Section~\ref{sec:intrinsic_stat}, but using the \texttt{KNN}-identified mergers instead of the intrinsic ones.
Figure~\ref{fig:stat_synthetic} presents the AGN and merger fractions in the same format as Figure~\ref{fig:stat_intrinsic}.
Compared to the intrinsic case, the AGN-merger connection is substantially weakened, particularly in the highest redshift bin ($1.5 < z < 3$), where the enhancement ratios approach unity.
While a modest AGN excess remains in the merger sample, the signal is significantly reduced relative to the intrinsic analysis.

Note that interpreting the merger fractions requires caution.
As discussed in Section~\ref{subsec:knn}, the \texttt{KNN} classifier tends to overpredict mergers due to a higher number of false positives compared to false negatives.
This leads to an overall increase in the merger fractions across all redshift bins, as seen in the upper panel.
For this reason, a more meaningful comparison is provided by the enhancement ratios ($R$ or $1/R$) in the lower panel.

The enhancement in merger fraction among AGNs ($R > 1$) persists in the mid- and low-redshift bins, indicating that some degree of AGN-merger connection is still recoverable using morphological classification alone.
However, the signal strength is notably diminished: in the low-redshift bin ($0.5 < z < 0.9$), the intrinsic analysis showed enhancement ratios of $R \sim 10$, whereas the mock-observation-based analysis yields only $R \sim 2$\--3.

These results demonstrate that limitations in observational merger identification, due to both morphological ambiguity and classification noise, can substantially weaken the detectable AGN-merger connection, even when the underlying physical link is strong.

As a sanity check against a commonly used human classification, we visually inspected a random subset (N=100) with equal numbers of \texttt{KNN}-classified mergers and non-mergers.
We found that the AGN incidence is higher in the consensus-merger bin than in the consensus non-merger bin, consistent with the \texttt{KNN}-based AGN–merger trend. Notably, the \texttt{KNN}-only merger subset (KNN=1, human=0) also shows an elevated AGN fraction relative to consensus non-mergers, suggesting that \texttt{KNN} can capture disturbed systems that may be visually ambiguous; overall, this qualitative agreement supports the use of our \texttt{KNN} selection for the full sample while still acknowledging residual classification scatter between methods.
A detailed investigation of the origin of the discrepancies between the two approaches is beyond the scope of this work.

\begin{table}
\centering
\caption{The exact values for Figure \ref{fig:stat_synthetic}.}
\resizebox{\linewidth}{!}{%
\begin{tabular}{lc|ccc}
\hline
                                                       & \multicolumn{1}{l|}{} & High-$z$      & Mid-$z$       & Low-$z$       \\
                                                       & \multicolumn{1}{l|}{} & ($3<z<1.5$)   & ($1.5<z<0.9$) & ($0.9<z<0.5$) \\ \hline
\multicolumn{1}{l|}{\multirow{2}{*}{$f_{\rm AGN}$}}    & merger                & 0.56$\pm$0.03 & 0.43$\pm$0.05 & 0.22$\pm$0.04 \\
\multicolumn{1}{l|}{}                                  & non-merger            & 0.57$\pm$0.04 & 0.35$\pm$0.06 & 0.07$\pm$0.04 \\ \hline
\multicolumn{1}{l|}{\multirow{2}{*}{$f_{\rm merger}$}} & AGN                   & 0.46$\pm$0.03 & 0.33$\pm$0.04 & 0.32$\pm$0.07 \\
\multicolumn{1}{l|}{}                                  & non-AGN                & 0.47$\pm$0.04 & 0.21$\pm$0.04 & 0.18$\pm$0.07 \\ \hline
\end{tabular}%
}
\label{tab:stat_synthetic}
\end{table}

\begin{figure*}
    \centering
    \includegraphics[width=0.85\linewidth]{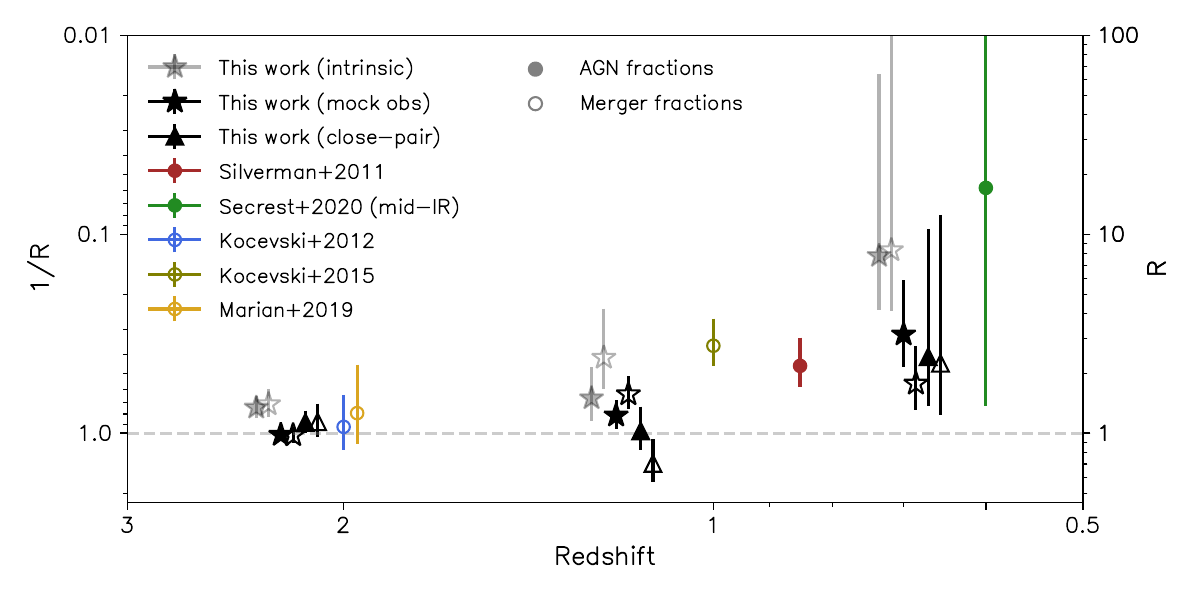}
    \caption{Collection of enhancement ratios ($R$ and $1/R$) from various observational studies, overlaid with the results from this work based on {\tt KNN}-identified mergers (black stars).
    The intrinsic results shown in Section \ref{sec:intrinsic_stat} are marked as grey stars.
    Additionally, the statistics using the close-pair definition ($<50\,\rm kpc$ separation with the mass ratio exceeding 0.25) are also shown as black triangles, which will be discussed in Section \ref{sec:mergerdefinitions}.
    The filled markers represent AGN fraction enhancements (i.e., AGN fraction in mergers vs. non-mergers), while open markers correspond to merger fraction enhancements (i.e., merger fraction in AGNs vs. non-AGNs).
    The left-hand $y$-axis shows $1/R$, the inverse enhancement ratio used in our analysis to mitigate issues when control sample fractions are low, while the right-hand $y$-axis shows the corresponding $R$ values for direct comparison with observational literature.}
    \label{fig:comparison_obs}
\end{figure*}

To conclude this section, Figure~\ref{fig:comparison_obs} presents a compilation of AGN/merger fraction excesses relative to control samples, based on various observational studies that provide well-defined comparison sets with statistical uncertainties, along with the results from this work for direct comparison.
\citet{Silverman2011} analyzed galaxies with $M_{\star} > 2.5 \times 10^{10}\,\rm M_{\odot}$ from the zCOSMOS 20K catalog, identifying AGNs as sources with Chandra-detected X-ray luminosities exceeding $2 \times 10^{42}\,\rm erg\,s^{-1}$. Kinematic pairs were selected based on projected separations of less than $100\,\rm kpc \,h^{-1}$ and line-of-sight velocity differences below $500\,\rm km\,s^{-1}$.
Both \citet{Kocevski2012} and \citet{Kocevski2015} classified merging galaxies using visual inspection and AGNs using X-ray luminosities, while the latter further identified heavily obscured AGN with $N_{\rm H}>10^{23.5}\,\rm cm^{-2}$ using X-ray spectral analysis.
Both \citet{Marian2019} and \citet{Secrest2020} visually identified mergers based on morphological distortions or tidal features. \citet{Marian2019} selected AGNs with high specific BH accretion rates (Eddington ratio > 0.7), while \citet{Secrest2020} adopted mid-IR AGN selection using the \textit{WISE} color cut of $W1-W2 > 0.5$. Although various AGN definitions were considered in the latter, we present only the one showing the highest enhancement.
For simplicity, we show only the median redshift for each observational study.

Our results, based on {\tt KNN}-identified mergers, are shown as black stars for comparison. Filled markers represent AGN fraction enhancements, while open markers indicate merger fraction enhancements, as noted in the figure caption.
This summary highlights how our mock-observation-based analysis is qualitatively consistent with observed trends, and illustrates the extent to which observational limitations may obscure the underlying AGN–merger connection.

\section{Discussion} \label{sec:discussion}
\subsection{Sensitivity to Merger Definitions: Mass Ratio, Timing, and Phase}
\label{sec:mergerdefinitions}
In the preceding sections, we adopted a fiducial merger criterion, in which a merger is defined by a stellar mass ratio greater than 0.25 and a post-coalescence time window of $\Delta t_{\rm merger} = 500\,\rm Myr$.
Here, we explore how variations in the merger definition affect the AGN–merger connection, focusing on three key components: (1) the stellar mass ratio threshold, (2) the inclusion of pre-merger phases, and (3) the width of the merger time window. This allows us to better understand the specific conditions under which mergers trigger AGN activity.

\begin{figure*}
    \centering
    \includegraphics[width=1.0\linewidth]{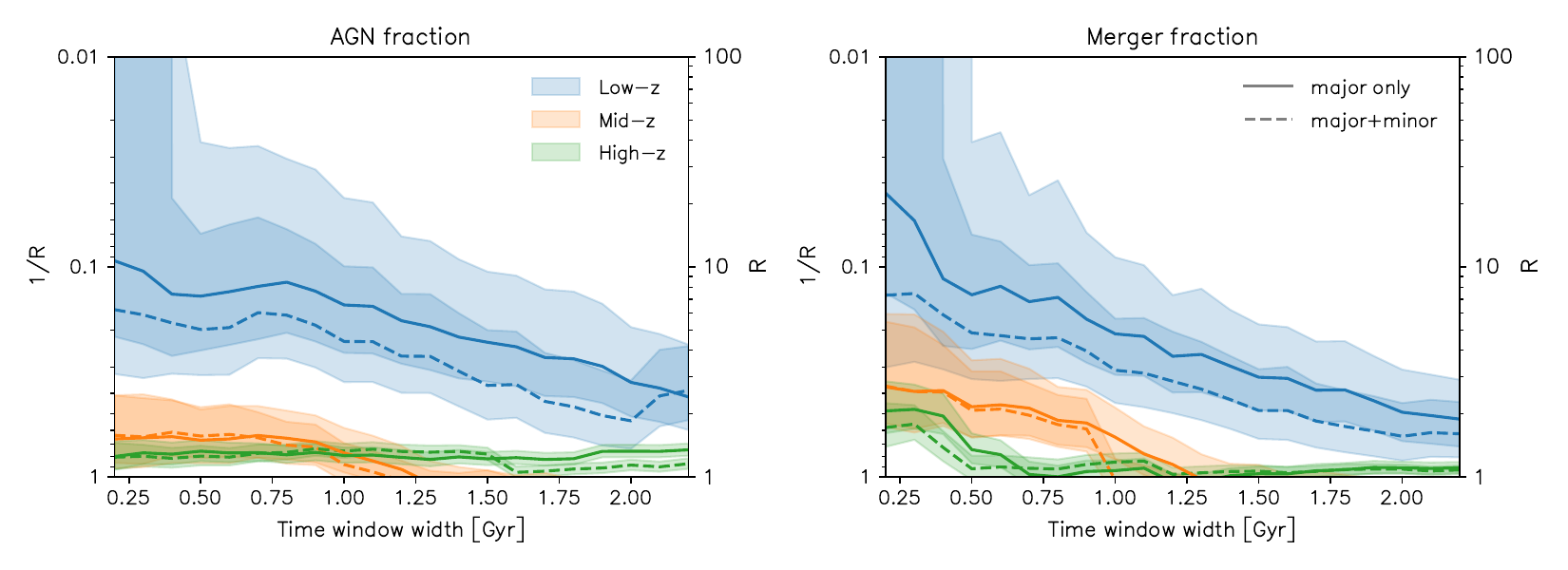}
    \caption{
    Impact of different post-merger definitions on the AGN–merger connection.
    The $x$-axis shows the width of the post-merger time window ($\Delta t_{\rm merger}$), while the $y$-axis shows $1/R$, the inverse enhancement ratio of AGN or merger fractions, as defined in Figures~\ref{fig:stat_intrinsic} and \ref{fig:stat_synthetic}.
    Solid lines correspond to definitions including only major mergers (stellar mass ratio $\geq 0.25$), while dashed lines include both major and minor mergers (stellar mass ratio $\geq 0.1$).
    Colours indicate redshift bins: low-$z$ (\textit{blue}), mid-$z$ (\textit{orange}), and high-$z$ (\textit{green}).}
    \label{fig:discuss_merger_defs_p}
\end{figure*}

Figure~\ref{fig:discuss_merger_defs_p} illustrates how the AGN-merger connection depends on mass ratio threshold and time window width, considering post-merger phases only.
The left panel shows the enhancement in AGN fraction among mergers, while the right shows the enhancement in merger fraction among AGNs.
Results are shown across three redshift bins: high-$z$ (green), mid-$z$ (orange), and low-$z$ (blue), with solid lines for major mergers ($\mu \geq 0.25$) and dashed lines for major+minor mergers ($\mu \geq 0.1$).

As seen in our previous analysis, the low-redshift bin consistently shows the strongest AGN-merger connection.
However, this signal weakens as the post-merger time window becomes wider.
For $\Delta t_{\rm merger} \gtrsim 2\,\rm Gyr$, the enhancement disappears entirely.
At higher redshifts ($z > 0.9$), the AGN fractions between mergers and non-mergers are largely indistinguishable across all time windows.
A slight enhancement is only visible for shorter time windows ($\lesssim 1\,\rm Gyr$) at intermediate redshifts and $\lesssim 0.5\,\rm Gyr$ at high-$z$.
The effect of mass ratio threshold is most notable in the low-$z$ bin, where major mergers consistently produce higher enhancement ratios than the inclusive (major+minor) case.
This supports our interpretation in Section~\ref{sec:intrinsic_stat} that more violent interactions are required to trigger AGN in gas-poor systems at low redshift.

In Figure~\ref{fig:discuss_merger_defs_prepost_lowz}, we isolate the role of the pre-merger phase in the low-$z$ bin by fixing the stellar mass ratio at 0.25 and comparing three definitions: (i) pre-merger only (dotted), (ii) pre+post-merger (dot-dashed), and (iii) post-merger only (solid).
The results clearly show that the post-merger phase contributes far more strongly to the AGN-merger connection than the pre-merger phase.
Both AGN and merger fraction enhancements are minimal when only the pre-merger phase is included.

This finding also helps reconcile the discrepancy between our results and those of \citet{Sharma2024}, who used the same simulation suite but defined mergers based on the presence of a nearby companion in single snapshots, effectively a pre-merger criterion.
In contrast, our use of full merger trees enables a more temporally integrated definition.
We also tried defining the galaxy mergers by checking the existence of close-pairs within $50\,\rm kpc$ from the central galaxy, and marked the fraction ratios as gray dashed lines in Figure \ref{fig:discuss_merger_defs_prepost_lowz}.
The relatively low AGN connection in the pre-merger phase and close-pair definition underscores the importance of precise and physically motivated merger definitions when investigating causal links with black hole activity.


\subsection{Comparison with Previous Works Using Simulations}
\label{sec:compare_prev}
In this subsection, we compare our results directly to several works that have also analyzed the AGN-merger connection using simulations: \citet{Sharma2024},  \citet{Byrne-Mamahit2024}, and \citet{Schechter2025}.

While \citet{Sharma2024} used the same suite of cosmological zoom-in simulations of massive galaxies as in this work, the key distinction lies in how mergers are identified.
They defined mergers based on close companion proximity in individual simulation snapshots, specifically, by locating stellar mass companions within a 50~kpc projected separation and above a 1:4 mass ratio threshold.
This effectively corresponds to a pre-merger phase definition that captures galaxies in interaction but not necessarily during or after coalescence.
In contrast, our approach leverages full halo merger trees, enabling a dynamic and physically motivated definition of mergers that encompasses the full temporal context of coalescence and subsequent evolution.
This allows us to define and vary the merger phase explicitly, distinguishing pre-merger, post-merger, or combined intervals, while also exploring how long the AGN-merger connection remains detectable.

In addition, while both studies use bolometric luminosity thresholds to identify AGNs, we further incorporate high-cadence black hole accretion histories.
This permits a more realistic AGN classification, based on whether a SMBH exceeded a given luminosity threshold at any point in the past $10^5$ years, mimicking observational time-delay effects such as narrow-line region response.

As a result of these methodological differences, some apparent discrepancies arise.
For instance, \citet{Sharma2024} report no significant AGN–merger connection at intermediate and high redshifts ($z>0.9$), and only a mild enhancement at low redshift ($0.5<z<0.9$), particularly in gas-rich mergers.
Our findings agree with this overall trend: we find that AGN activity is indeed more tightly coupled to mergers at lower redshifts, where galaxies tend to be more gas-poor and dynamical events are more effective in fueling the central SMBH.
Moreover, our sensitivity tests in Section~\ref{sec:mergerdefinitions} show that restricting to pre-merger phases does significantly weaken the apparent AGN–merger connection as in \citet{Sharma2024}.

\citet{Byrne-Mamahit2024} analyzed the correlation between post-mergers and AGN using galaxies from the IllustrisTNG simulation at $z<1$.
Similar to our approach, they defined merger events from merger trees and tested the connection by comparing the AGN (or merger) fraction ratios between a test sample and a control sample.

While their redshift range corresponds to our lowest redshift bin ($0.5<z<0.9$), our simulation is biased toward more massive galaxies, as we zoomed in on massive galaxies with a stellar mass of $\gtrsim 10^{11} \rm\,M_{\odot}$ at $z=0$.
This characteristic, as mentioned in Section \ref{sec:intrinsic_stat}, has an impact on the gas fraction distribution in the central halo.
Consequently, understanding the causal link between AGN and galaxy mergers in our study is effectively equivalent to understanding it as a function of gas fraction.
With this in mind, the finding by \citet{Byrne-Mamahit2024} that the ratio of the AGN (merger) fraction between their test and control samples is at most $\lesssim 2$ for the post-merger + moderate AGN ($L_{\rm bol}\gtrsim10^{44}\,\rm erg/s$) combination can be considered consistent with our results.

\begin{figure*}
    \centering
    \includegraphics[width=1.0\linewidth]{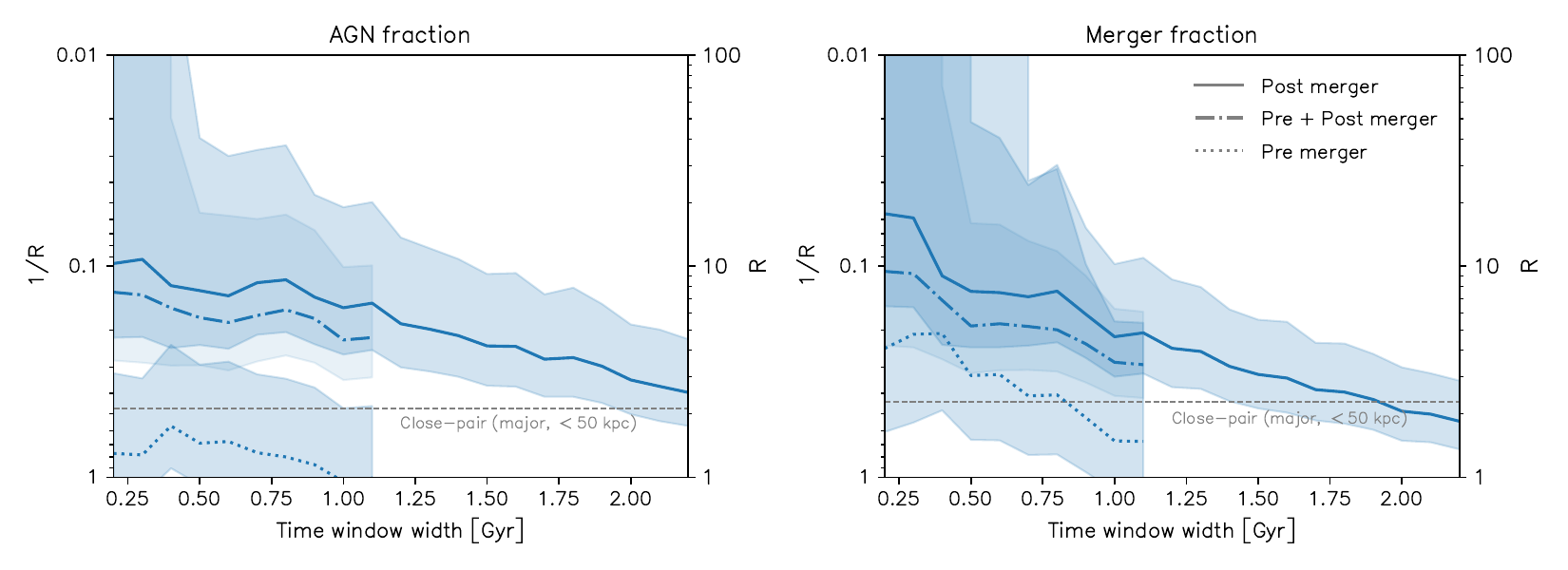}
    \caption{
    Comparison of AGN–merger connectivity for different merger phase definitions in the low-redshift bin. $x$- and $y$- axes are the same as in Figure \ref{fig:discuss_merger_defs_p}. Solid lines correspond to the post-major-merger definition, identical to the blue solid line in Figure~\ref{fig:discuss_merger_defs_p}. Dotted lines represent pre-merger-only definitions, while dot-dashed lines include both pre- and post-merger phases. For definitions involving pre-merger phases, the time window is limited to $\lesssim 1.1 \,\rm Gyr$ due to the requirement of stable identification of halo progenitors. Additionally, the grey dashed lines represent the fractions when galaxy mergers are identified based on the presence of close pairs with the stellar mass ratio exceeding $0.25$ within $50 \,\rm kpc$. The associated uncertainties ($\sim 0.35$ and $\sim 0.34$ in each panel) are omitted in the figure for clarity.
    }  \label{fig:discuss_merger_defs_prepost_lowz}
\end{figure*}

Furthermore, their study suggests that the AGN–merger connection is quantitatively similar between the post-merger and pre-merger phases, which contrasts with our findings, where we find the pre-merger phase to be less significant than the post-merger phase.
However, Figure \ref{fig:discuss_merger_defs_prepost_lowz} explains the relative importance of the post-merger phase over the pre-merger phase for low-redshift galaxies with low gas fractions.
We found that as the gas fraction increases (which corresponds to higher redshifts), the relative importance of the pre-merger and post-merger phases becomes increasingly similar.
This suggests that matching the galaxy's gas fraction is crucial for a meaningful comparison of the AGN-merger connection across different studies.

\citet{Schechter2025} analyzed galaxy mergers in the IllustrisTNG50 simulations over $0.2<z<3$, a redshift interval closely matching that explored here.
Instead of focusing on an excess in the AGN fraction, they quantified merger-driven activity by measuring the enhancement in the specific black-hole accretion rate (sBHAR) of merging systems relative to a matched non-merging control sample.
They found that the merger-associated sBHAR enhancement strengthens toward higher black-hole masses, and towards lower redshift, which they interpreted as being connected with the gas fraction in the merger progenitors. 
This is consistent with our finding that the gas fraction plays a key role in shaping the statistical connection between mergers and nuclear activity.
A particularly noteworthy aspect of their results is that, at the highest redshifts, major mergers do not necessarily yield an enhancement; instead, the measured excess can fall below unity, indicating that merging galaxies may accrete less than their non-merging counterparts.
To interpret such behavior, \citet{Schechter2025} discuss how the dependence of merger-driven accretion on the pre-merger gas supply can be non-monotonic: in very gas-rich galaxies, substantial star formation during the pre-coalescence phase may consume or redistribute the gas in ways that diminish a distinct merger-triggered sBHAR boost, while in very gas-poor galaxies the scarcity of fuel can likewise suppress any enhancement.
We similarly find that the gas content immediately prior to coalescence is an important parameter for understanding the statistical AGN–merger association.
In our case, however, we do not observe a tendency for the excess to drop below unity at the highest redshifts, plausibly because our sample selection (massive ellipticals at $z=0$) biases the high-redshift progenitors toward higher stellar masses at fixed epoch.

In summary, while the methods differ, most notably in how mergers are defined, in the temporal resolution of AGN classification and in the target galaxies, the conclusions of these studies are in broad agreement:
major mergers are not the most dominant driver of AGN activity across cosmic time, but can play a significant triggering role, especially in gas-poor galaxies.

\section{Summary and Conclusions} \label{sec:summary}
Using a suite of cosmological zoom-in simulations of 31  galaxies, we investigate the connection between galaxy mergers and AGN activity, a topic where observational studies have produced conflicting results.
We define mergers based on merger trees constructed from six-dimensional phase-space information, and identify AGNs using detailed black hole accretion histories.
To assess the observational detectability of this connection, we generate realistic mock JWST images and analyze their morphological properties.
Our main findings are as follows:
\begin{enumerate}
    \item A clear statistical association between galaxy mergers and AGN activity is found at low redshift ($0.5 < z < 0.9$), where AGN fractions among mergers are significantly enhanced.
    This is not necessarily because mergers only trigger AGN at low redshift, but rather because central gas reservoirs are more depleted in this regime, making mergers more critical for channeling gas into the SMBH.
    \item The AGN triggering is strongest in major mergers (stellar mass ratio $\geq 0.25$) and during the post-merger phase within $\sim$0.5-1 Gyr after coalescence.  This supports the scenario in which violent merger-driven dynamics effectively funnel gas into the nuclear region after a delay. In contrast, pre-merger phases show little to no enhancement in AGN activity.
    \item Rather than relying on empirical cuts in two-dimensional morphological parameter space, we employ a $k$-nearest neighbours (KNN) algorithm in a five-dimensional space, comprising $G$, $M_{20}$, $A$, $C$, and redshift, which significantly improves merger identification from mock observations.
    \item Observational limitations introduce significant uncertainty in detecting AGN-merger connections.  
    Our synthetic analysis shows that the intrinsic signal is substantially weakened once realistic observational constraints are applied: at high redshift ($z > 1.5$), the AGN–merger connection effectively disappears, and even at low redshift, the enhancement ratio $R \sim 10$ is reduced to $R \sim 2$-3.  
    This result offers one potential explanation for the wide variation in observational findings regarding the AGN-merger connection.
\end{enumerate}

\section{Acknowledgements}
    This work was supported by the 2024 Research Fund of the University of Seoul for E.C. T.N. acknowledges the support of the Deutsche Forschungsgemeinschaft (DFG, German Research Foundation) under Germany's Excellence Strategy - EXC-2094 - 390783311 of the DFG Cluster of Excellence ``ORIGINS''. JK acknowledges the support of the National Research Foundation of Korea (NRF) grant funded by the Korea government (MSIT) (2022M3K3A1093827). JK also acknowleges the support of the Center for Advanced Computation at Korea Institute for Advanced Study.

    This work makes use of the following software packages: \texttt{Powderday} \citep{Narayanan2021}, \texttt{yt} \citep{Turk2011}, \texttt{Hyperion} \citep{Robitaille2011}, \texttt{FSPS} \citep{Conroy2010}, \texttt{statmorph} \citep{Bottrell2019}, \texttt{Scikit-learn} \citep{Pedregosa2011}, \texttt{Imbalanced-learn} \citep{Lemaitre2017}, and \texttt{Astropy} \citep{AstropyCollaboration2013, AstropyCollaboration2018}. \texttt{Powderday} was conceived of at the Aspen Center for Physics. The Aspen Center for Physics is supported by National Science Foundation grant PHY-2210452. The Flatiron Institute is supported by the Simons Foundation.


\section{Data Availability}
The data underlying this article will be shared on reasonable request to the corresponding author.



\bibliographystyle{mnras}
\bibliography{library} 


\appendix
\label{sec:appendix}

\section{Parameter Selection for {\tt KNN} Merger Classifier with Oversampling}
\label{sec:_optimal_params}

In Section~\ref{subsec:knn}, we adopted a $k$-nearest neighbours (KNN) classifier in combination with random oversampling to address the strong class imbalance in our dataset.
Here, we describe how we selected the optimal combination of two key parameters: the number of nearest neighbours $k$ used in classification, and the oversampling ratio, denoted by {\tt sampling\_strategy}, which defines the final number of oversampled minority (merger) samples relative to the number of majority (non-merger) samples.

Because only about 10\% of the total test set consists of mergers, naively maximizing overall accuracy would favor trivial classifiers that simply predict the majority class.
For instance, labeling all samples as non-mergers yields an accuracy of 0.9, despite having no predictive power for mergers.
Therefore, it is crucial to adopt evaluation metrics that are appropriate for imbalanced datasets.

We tested several such metrics, including the balanced accuracy (BA) and the geometric mean (GM) of the true positive rate (TPR) and true negative rate (TNR), both of which account for performance across both classes:
\begin{equation}
    \begin{aligned}
        & \rm BA = \frac{TPR+TNR}{2} \\
        & \rm GM = (TPR\cdot TNR)^{1/2} \\
    \end{aligned}
\end{equation}

\begin{figure*}
    \centering
    \includegraphics[width=1.0\linewidth]{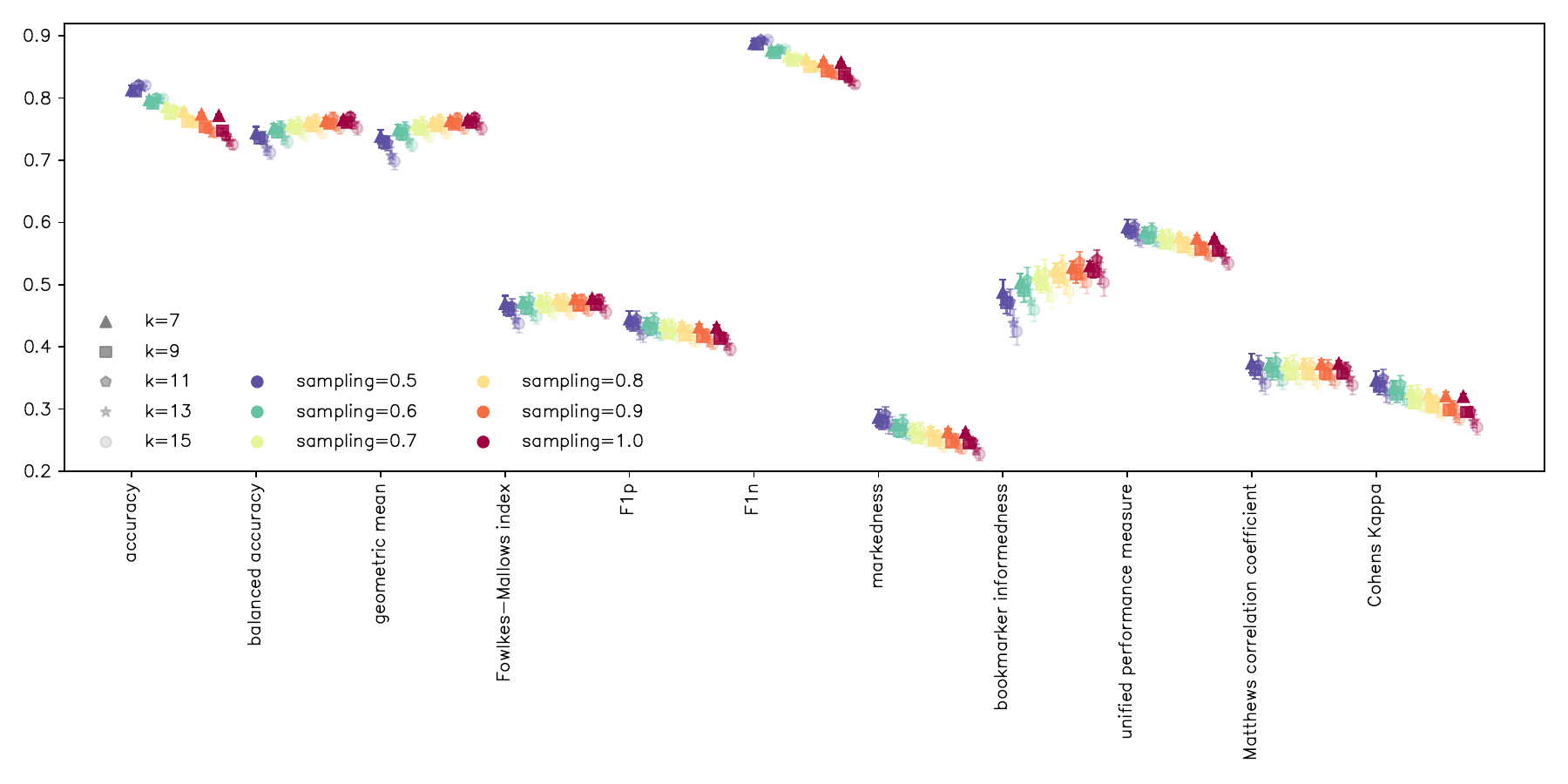}
    \caption{
    Performance of different classification metrics for varying combinations of {\tt KNN} model parameters.
    Colours represent the oversampling ratio {\tt sampling\_strategy}, ranging from 0.5 (\textit{blue}) to 1.0 (\textit{red}), where a value of 0.5 corresponds to oversampling the minority class to half the size of the majority class.
    Transparency levels indicate the number of nearest neighbours $k$ used in the classifier.
    All models use $p=1$ (Manhattan distance) for distance measurement.
    Error bars denote the standard deviation over 100 random oversampling realizations.
    }
    \label{fig:exploring_parameter_set}
\end{figure*}

Figure~\ref{fig:exploring_parameter_set} presents 11 classification metrics evaluated over a grid of $k \in {5,7,9,11,13}$ and {\tt sampling\_strategy} $\in {0.5, 0.6, 0.7, 0.8, 0.9, 1.0}$.
Each combination was repeated 100 times with random oversampling, and the shaded error bars indicate the standard deviation across these realizations.

We find that overall accuracy decreases with increasing oversampling and larger $k$, whereas balanced accuracy and geometric mean increase with oversampling and peak around $k=11$.
While the differences between $k=9$ and $k=11$ are minor, we chose {\tt sampling\_strategy} = 1.0 and $k=11$ as our fiducial configuration based on the balanced accuracy.
We also verified that our main scientific conclusions are not sensitive to modest variations in these parameters.

\bsp	
\label{lastpage}

\end{document}